# Practical LTE and Wi-Fi Coexistence Techniques beyond LBT

Jonathan Ling, David Lopez-Perez, Mohammad R. Khawer

ABSTRACT

Coexistence with Wi-Fi is the key issue for unlicensed band LTE. The main coexistence mechanism is Listen-Before-Talk, whereby radio frequency energy is sensed over a short period of time and compared to a threshold. Given the default energy thresholds, the energy sensing range is actually much less than the cell range. Both technologies can experience collisions due to transmission being below energy detection threshold. Currently Wi-Fi is agnostic of LTE presence in the unlicensed spectrum. To improve coexistence a communications channel via relaying is proposed to be used by the unlicensed band LTE, to announce its presence on an unlicensed channel. Legacy Wi-Fi APs may be programmed to interpret and respond by firmware upgrade at the AP to enhance its channel selection algorithm. Higher performance for both networks is demonstrated via more effective radio frequency channel selection and adaptive energy detection thresholding.

## I. INTRODUCTION

Fair coexistence between wireless technologies in the unlicensed band in a distributed manner for most locations and time is challenging [1][2]. To coordinate inter-technology spectrum access in a distributed and simple manner, a transmitter must first detect the energy across the intended transmission band. This energy detection (ED) mechanism informs the transmitter of ongoing transmissions by other nodes, and helps it to decide whether to transmit or not. However, although simple, this scheme, also known as listen-before-talk (LBT), does not work in all circumstances, for example, when information is encoded to be received below background noise level, or when the nodes are distant and the signals are weak at the receiver. Thus, a node wishing to transmit may sense the channel as unoccupied according to the energy received being below a certain ED threshold, but may still interfere with a nearby node that is receiving. Nonetheless, LBT is the starting point for coexistence, and is mandatory in many countries' unlicensed band regulations. The ED threshold cannot be lowered too much as false detections would occur due to noise. Consequently, there is a need for additional information for effective inter and intra technology wireless media access.

The 802.11 media access control (MAC) protocol augments the ED mechanism with a virtual carrier sense (VCS) mechanism, whereby 802.11 packet headers are received and decoded at the lowest power levels due to using the most robust modulation and coding [3]. The network allocation vector (NAV), i.e. timeline at each station (STA) of when the channel is free or occupied, is updated based on the contents of such header or control packets, which indicate for how long the channel will be used. For example, the request-to-send/clear-to-send (RTS/CTS) mechanism reserves the channel by causing the NAV to be updated by all nodes that receive the RTS around the transmitter and CTS around the receiver. However, even the VCS has problems, as the capture effect, which causes the stronger overlapped packet to be captured over the weaker one, results in unfairness as the stronger node doesn't experience a collision and the weaker node backs off.

The term $u$LTE for unlicensed long term evolution (LTE) will be used to denote all unlicensed band LTE specifications: LTE-Unlicensed (LTE-U), Rel-13 licensed assisted access (LAA), Rel-14 enhanced LAA (eLAA), and MulteFire. $u$LTE MAC protocols go beyond the regulatory requirements with improved fairness to wireless-fidelity (Wi-Fi) signals. For example its ED threshold at -72 dBm is 10 dB lower than Wi-Fi. Moreover, third generation partnership project (3GPP)'s CAT4 LBT has an exponential back off mechanism similar to 802.11, and is designed with the awareness of the 802.11 MAC [4]. Still, this process is based on energy detection, and due to the lack of the 802.11 VCS, it may exhibit unfairness when Wi-Fi signals fall below the ED threshold.

In the dynamic spectrum sensing community, explicit inter-radio-access-technology (RAT) signaling is used to coordinate channel access. In some solutions, secondary clients of the band are allowed to use a channel after a database lookup and registration. However, such approach is not possible indoors due to limited availability of accurate position. Another approach is to design from the ground up a new common MAC protocol for both $u$LTE and Wi-Fi, but that would ignore the huge installed base of 802.11 stations. Many solutions have been proposed to have the $u$LTE radio working in conjunction with a Wi-Fi radio, including helping to adapt $u$LTE transmission period based on Wi-Fi traffic sensing [4] and Wi-Fi headers on $u$LTE transmissions [6]. Both schemes have demonstrated benefits, but the neighboring Wi-Fi nodes are not explicitly aware of $u$LTE. This has the drawback that Wi-Fi nodes are unable to adapt accordingly, e.g. during channel selection, which can be especially important for the uplink reception at the eNB.

In this article, we propose a novel signaling framework that allows each technology to be aware of the presence of the other. Moreover, we propose to use such information to allow an enhanced channel selection as well as an adaptive ED threshold tuning. The remainder of this article is organized as follows: Section II discusses the limitations of current

Jonathan Ling (Jonathan.Ling@nokia.com) and David Lopez-Perez (David.Lopez-Perez@nokia.com) are with Nokia Bell Labs. Mohammad R. Khawer (Mohammad.khawer@nokia.com) is with Nokia Mobile Networks CTO.

signaling mechanisms and proposes two relaying techniques. Section III describes how channel selection is improved using this relaying channel. Section IV discusses some limitations of LBT and ED due to radio propagation and its effect on a protocol. Section V, motivated by these results, shows how the MAC can be enhanced and ED thresholds tuned to improve fairness. Finally, the Lessons Learned section summarizes the work and provides thoughts on future directions.

## II. CROSS TECHNOLOGY COMMUNICATIONS CHANNELS

Basic coordination of spectrum resources requires that each system identify itself, e.g. type of physical and MAC layer and other network features. Based on the neighboring cell identification, i.e. Wi-Fi access points (AP) or LTE enhanced NodeBs (eNB), a base may first select the "cleanest" channel and secondly adjust the politeness of its MAC. Self-identification is also necessary to further improve the self-organizing-network capabilities (SON). Cell discovery mechanisms in Wi-Fi and LTE and their issues are discussed in the section next, followed by a discussion on practical inter-RAT communication channels.

### A. Cell identification

Most readers are familiar with the Wi-Fi beacon, which is a message broadcast to all Wi-Fi Stations (STAs) describing various characteristics of the Wi-Fi Access Point (AP), such as its service set identifier (SSID), channel, timestamp and the features it supports. The beacons are repeated regularly, encoded in at the lowest Modulation & Coding Scheme (MCS), and access the channel as a priority frame according to the regular distributed coordination function (DCF) procedure.

In contrast, cell discovery in LTE begins by the user equipment (UE) attempting to decode the physical broadcast channel (PBCH). Since UE attachment is network directed, the PBCH contains only the information that is needed to build a connection. The Master Information Block (MIB) contains the system bandwidth and the system frame number, and is repeated every 40 ms. The MIB is detected via autocorrelation of the primary synchronization sequence (PSS). Additional system information in the system information blocks (SIBs) is carried on the physical downlink shared channel (PDSCH), which is time multiplexed over the 40 ms slots. SIB1 contains the operator identifier (PLMN) and the cell identifier, among others.

There are two types of LTE access on unlicensed frequencies: LAA, which acts as a supplemental downlink to a licensed LTE carrier (note: unlicensed uplink eLAA is still attached to licensed carrier), and MulteFire, which is characterized by fully standalone operation in the unlicensed band. In LAA, both the licensed and unlicensed bands are operational at the same time, i.e. data may be received over both bands simultaneously. The PBCH is carried only on the licensed carrier. However Rel-12 discovery reference signals (DRS), which includes the PSS, are transmitted at 40 ms intervals on the unlicensed carrier. Detection of DRS alone does not provide further information, i.e. cell id, and one cannot even determine the operator. MulteFire transmissions do include the PBCH/PDSCH in their downlink transmissions, now called ePBCH, which doubles the energy in the PSS and secondary synchronization signal (SSS) sequences to improve detect-ability.

Notably Wi-Fi through the 802.11aa specification shares additional information, such as cell loading per traffic characteristic, to support improved coexistence in so called "overlapped BSS" scenarios. $u$LTE doesn't yet have such ability, but an "LTE "beacon that provides cell loading has been proposed [7].

### B. Network and client based relaying

For inter-RAT cell identification, one might include a LTE receiver in every Wi-Fi AP that could decode PBCH and PDSCH. Note that both MulteFire and LAA would need a new SIB or "LTE beacon" for inter-cell coordination. Likewise, every LTE receiver could include a Wi-Fi receiver that can decode Wi-Fi beacons and management frames. Clearly this imposes additional costs and requirements. A better solution would be an approach that utilizes those Wi-Fi and LTE receivers that are naturally co-located together, e.g. at UEs. Furthermore the approach should be compatible with legacy Wi-Fi APs, in the sense of requiring at the most a software upgrade.

This UE based approach may be taken when the AP or eNB has a smart phone with both Wi-Fi and unlicensed band LTE modems, as shown in Figure 1a. This permits an apparently straightforward solution whereby the LTE modem asks for a channel scan from the Wi-Fi modem, and vice versa. This UE assisted scheme works best with enterprise Wi-Fi networks that utilize such measurements in their SON algorithms. In contrast, home Wi-Fi networks are less likely to utilize these measurements. One disadvantage of the UE based approach is that a dual RAT UE must be available, and for optimal performance, regular scans must be taken, reducing battery life. Moreover, multiple UE may be necessary to detect all the neighboring cells.

In order to solve the mentioned UE based approach issues, we propose a network based approach, in which a $u$LTE base communicates directly to a "friendly" Wi-Fi AP. These friendly Wi-Fi APs may be utilized to relay LTE system information and loading to surrounding Wi-Fi networks via the Wi-Fi beacon mechanism that we refer to as the Pseudo Beacon for $u$LTE small cells, and vice versa, as shown in Figure 1b.

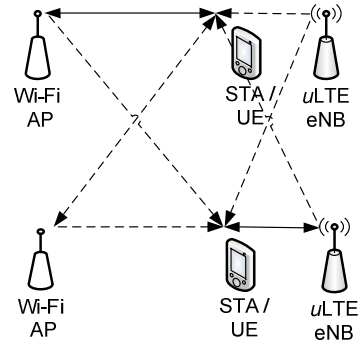

*(a) UE-based approach*

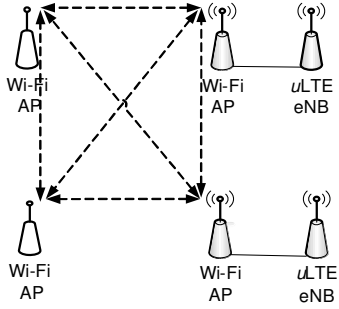

*(b) Network-based approach*

Figure 1. Inter-RAT relaying via client (a) and via the base station (b). Solid lines denote relaying channel. Dashed lines denote over the air detection.

### III. ENHANCED CHANNEL SELECTION:

Thanks to the proposed network based approach to advertise the presence of one technology to the other through a relay node, Wi-Fi can take into account $u$LTE while performing channel selection. In more detail, the helper Wi-Fi AP would transmit a Wi-Fi beacon with the ID of the $u$LTE cell and other information such as loading, as shown in Table 2. This Pseudo Beacon makes the $u$LTE cell appear as another Wi-Fi AP to unmodified APs and STAs, providing partial backward compatibility. However the full interpretation of the Pseudo Beacon requires a firmware upgrade at the AP. The information of Table 2 can be provided to channel selection and adaptive ED thresholding algorithms to improve the overall network co-existence and performance. Whereas before a Wi-Fi AP was not aware of the presence of an $u$LTE eNB now it is.

The goal of channel selection algorithms is to provide the highest performance for its own UEs in the long term. Alternatively, it may be expressed as choosing the channel with the least interference or least activity. Channel selection is performed infrequently, as scanning it prevents UE traffic from being served. Some UEs may also incur service interruption if they are unable to interpret channel switch announcements. For the purpose of expositional, consider the following channel selection algorithm run on a Wi-Fi AP:

1. Let $S$ be result from scan of AP/eNBs
2. Let $S_f$ be the filtered $S$ to remove weak bases according to the RSSI being less than threshold.
    - For $u$LTE eNBs, adjust RSSI according to the "TX power offset" field of Table 2.
3. Compute the channel usage based on the combined metric of predicted and actual airtime usage: $M_c = w_1*\text{average}(U) + w_2*\text{sum}(N_{\text{Attached}})$
    - For $u$LTE eNBs, according to the "MAC Spec" field of Table 2, adjust upwards the partial metric to account for difficulty in timesharing.
4. Select the channel with the minimum metric: $C^* = \text{argmin}(M)$.

where $U$ is a list of utilizations of $S_f$ and $N_{\text{Attached}}$ is the number of attached clients of $S_f$.

Step 2 filters weak APs, as their transmissions will tend to overlap and reuse the channel despite the virtual carrier sense. Step 3, computes a metric $M$, which attempts to balance current usage with future usage and how the channel is time shared according to the weights $w_1$ and $w_2$.

Based on the above, we observe how relaying enables the channel selection algorithm to correctly detect the base stations on each channel, whether Wi-Fi or $u$LTE, and devise a metric based on their reported and predicted utilization. Fine adjustment is made based on a node's MAC classification. The selection algorithm, when running on the eNB, would adjust upwards the metric in step 3 to account for difficulty in timesharing with Wi-Fi, causing the eNB to preferentially contend with other eNBs.

### IV. LIMITATIONS OF LBT

The effect of short and large scale propagation on channel sensing is discussed first, followed by its impact on the 802.11 DCF and a leading $u$LTE MAC protocol.

#### A. Radio Propagation

The wireless channel is characterized by both small and large scale fading. Small scale fading on the order of wavelengths is due to multipath, while large scale fading on the order of 10s of wavelengths is due to features of the environment. Large scale fading also has a random component referred to as random shadow fading, which is the effect produced by materials such as wood and concrete, both absorbing and scattering radio energy.

Proper system coordination through ED requires *all* stations to receive *all* transmitted signals above the ED threshold *all* the time. This is a difficult requirement to satisfy,

$$P\{ED\ success\} = \prod P\{P_t G(n,m) > ED\ threshold\}$$

where $P_t$ is the transmit power, and $G(n,m)$ is the path gain from node $n$ to node $m$, considering both short and large scale fading.

In order to verify such difficulty, let us assume there are 5 nodes wanting to access the channel, i.e., there are 10 links, counting reciprocal links as a single link, and that short scale fading power is Chi-square distributed. Notably 10% of the time there is a 10 dB or greater fade. Given the above fading channel, an ED threshold of -62 dBm, and all nodes receiving each other at -52dB on average, $P\{ED\ success\} = (.90)^{10} = 34\%$. This shows that sensing errors are prevalent due to small scale fading even for a small number of nodes and when the average signal strength is much higher than the ED threshold.

As another example of the occurrence of bellow ED threshold conditions, let us now assume a simulated building of 50 X 120 meters, a single base station that is located off center at (25m,30m), according to 3GPP 36.814, and UEs distributed uniformly, following a Poisson point process. Moreover, assume Wi-Fi and the $u$LTE base stations are place next to each other, i.e. the path gain $G$ from the base stations to its UEs is approximately the same as that among the AP and the eNB, and that the AP and eNB have a single client each.

Building type sensitivity is investigated for the above setup considering i) a "open" building using the InH propagation model given in TS 36.814, and ii) a "light partitioned" building using a diffusion model with its parameters defined in [8]. The diffusion model correctly represents signal strength at large distances, whereas the error of single slope models increases with it. In this study, only the average signal strength is considered, and an additional margin is needed to account for the multipath fading described previously.

Figure 2 provides the cumulative density function of the received signal strength (RSSI) for 20 dBm transmit power, and for both propagation models. Vertical lines show the minimum Wi-Fi signal strength of -87.5 dBm, for MCS0, according to Table 5.2 of [3], and the minimum LTE signal strength of -100 dBm, for QPSK R=1/8, assuming 6 dB UE noise figure according to Chapter 21.4 of [9]. ED thresholds are also highlighted at -62 dBm and -72dBm for Wi-Fi and $\mu$LTE, respectively. Applying the minimum signal strength threshold, Wi-Fi cell coverage is 87% by InH and 62% by diffusion. Likewise, $u$LTE coverage is 100% by InH and 75% by diffusion, which reveals the impact of the large scale fading.

From Figure 2, clearly a large portion of the cell coverage area receives power less than the ED threshold. Let the fractional ED coverage be defined as the fraction of the cell area that receives signal greater or equal to the ED threshold. Table 1, provides the fractional ED coverage for cell type, threshold, and building type.

Now consider uplink reception at the bases, and assume due to antenna gain at the bases the equivalent EIRP is 20 dB, i.e. the same as the downlink. Sensing failure of the other technology's transmission is given by

P{*ED failure at uLTE*}=
   P{ RSSI < -72 dBm at eNB | STA TX } and
P{*ED failure at Wi-Fi*}=
   P{ RSSI < -62 dBm at AP | UE TX }.

For this scenario P{*ED failure at Wi-Fi*} = 1-P{ RSSI >-62 dBm at AP}. Using Table 1, column 2, P{*ED failure at Wi-Fi*}=55% for InH and 74% for the diffusion model. Likewise, using Table 1, column 3, P{*ED failure at uLTE*}= 42% for InH and 33% for the diffusion model. This shows that there is a large area where signals are received below ED threshold, and inter-technology coordination is not possible relying on ED alone.

*B. LBT Protocol relying on ED*

Downlink collisions can still occur even when the AP and a eNB receive each other's transmissions at level much higher than their respective ED thresholds. For example, let the eNB follow the Cat 4 specification in 36.213, whereby there is a defer period of at least 1 SIFS + 1 slot. That is, the eNB is aware of the 802.11 protocol where there is data transmission followed by a short inter-frame space (SIFS), followed by the acknowledgement (ACK). On downlink transmission of Wi-Fi, assume the eNB easily detects the energy, and refrains from transmitting, as shown in Figure 3a. After the end of the data packet, the energy drops, the channel is clear, and the eNB starts a timer for 1 SIFS + 1 slot. If the ACK is detected, i.e. if it is received at greater than -72 dBm, the eNB will refrain from transmitting. If the ACK is received below -72 dB, then the eNB will transmit, as shown in Figure 3b. It is then clear that collisions are possible whereby an ACK is in the process of being received, while the eNB does not detect it, and goes on to transmit, as shown in Figure 3c. The ED threshold of -72 dB, effectively limits the collision-free downlink range of the AP, i.e. the Wi-Fi UE must be close enough to the AP and eNB such that the ACK is received at -72 or greater.

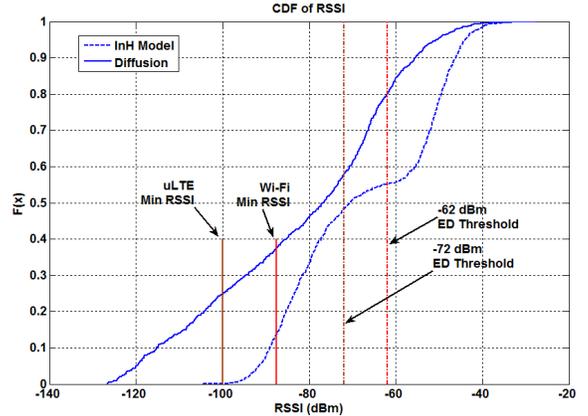

Figure 2. LBT coverage given two indoor propagation models and TX power of +20dBm.

|  | -62 dBm Threshold | | -72 dBm Threshold | |
| --- | --- | --- | --- | --- |
|  | Wi-Fi Cell | uLTE Cell | Wi-Fi Cell | uLTE Cell |
| InH | 51% | 45% | 58% | 52% |
| Diffusion | 32% | 26% | 67% | 56% |

Table 1. Fraction of cell area ED is active for given {cell type, threshold, propagation model}.

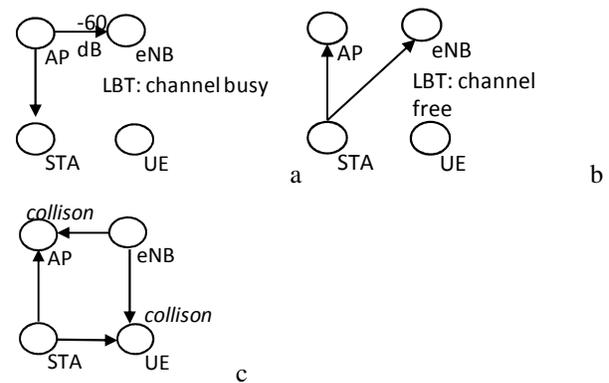

Figure 3. Collision Scenario: (a) Downlink transmission (solid line) received above ED, (b) uplink ACK (dashed) received below ED (c) collision at AP and UE due to transmitting eNB.

V. ADAPTIVE ED THRESHOLDING

After the channel selection process is complete, we propose that the ED thresholds be adapted to improve efficiency and fairness when both Wi-Fi and $u$LTE share the same channel. Consider the following algorithm to be run on a base station.
1. Initialize $T=T_{default}$.
2. Let $S$ be result from scan of co-channel AP/eNBs
3. Let $S_f$ be the filtered $S$ with clients greater than 0.
4. Set $T=$ min( $\underline{T}_{min}$, $R_f$ )
5. After delay $\tau$ goto 1.

where $R_f$ a vector of RSSIs associated with $N_f$, and $\tau$ controls the update rate.

We simulated network performance a Wi-Fi cell coexisting with a $u$LTE cell, according to the scenario in Figure 4. The RSSI from the Wi-Fi AP to the $u$LTE falls below the default ED threshold and vice versa. Figure 5 shows the results in terms of UE/STA downlink file throughput. Without the adaptive ED scheme, the Wi-Fi AP and the $u$LTE eNB do not detect each other, and thus their downlink transmissions are not coordinated. This results in a high number of collisions and re-transmissions, which mostly affect Wi-Fi performance. This is because Wi-Fi quickly detects the collision through the RTS/CTS mechanism and continuously backs-off, while $u$LTE, which does not have RTS/CTS, continues to transmit effectively forcing Wi-Fi off of the band. Eventually, the Wi-Fi AP transmits to its STA, when the $u$LTE eNB leaves the band empty not having any data to transmit to its UE. In contrast, with the adaptive ED scheme, the Wi-Fi AP and the $u$LTE eNB sense each other's transmissions, thus they are able to coordinate. This results in fairer time sharing, as well as much fewer collisions and retransmissions. Such coordination benefits Wi-Fi as it does not back off as much to $u$LTE, but necessarily impacts $u$LTE as it decreases its air time to share with Wi-Fi.

In terms of median throughput, Wi-Fi performance increases 3.5x (from 15 Mbps to 54 Mbps), while $u$LTE decreases 36% (from 49 Mbps to 31 Mbps). It is important to note that the overall system performance increases with the proposed adaptive ED threshold scheme by 32% (from 64 Mbps to 85 Mbps). In short, with adaptive thresholding, the channel is shared more fairly and the overall efficiency has improved.

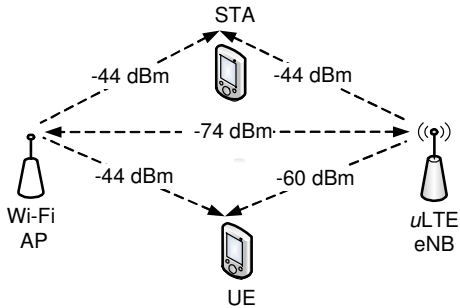

Figure 4: Scenario with Wi-Fi and $u$LTE each with a single client.

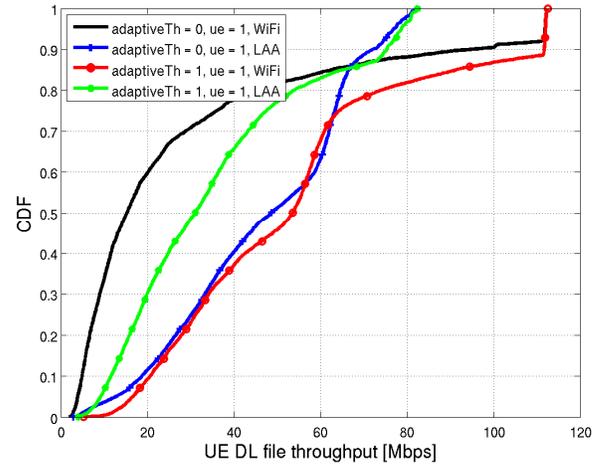

Figure 5: Downlink file throughputs for scenario in Figure 4.

### LESSONS LEARNED

Below ED threshold signals occur for a large fraction of the cell area, which demonstrates that inter-technology (Wi-Fi, and $u$LTE) coordination is not possible by relying on ED alone. A relaying channel based on co-located modems is proposed to facilitate cross technology communications. With network based relaying, Wi-Fi beacons may be used to identify unlicensed band LTE cells. This Pseudo Beacon makes the $u$LTE cell appear as another Wi-Fi AP to unmodified APs and STAs, providing partial backward compatibility. However the full interpretation of the information requires a firmware upgrade at the AP. This may aid in channel selection and in adapting MAC parameters as both technologies may now positively identify each other's presence on an unlicensed channel. Simulations show improvement in throughput when both technologies (Wi-Fi, and $u$LTE) adapt their ED thresholds and coordinate with each other.

As usage of both LTE and Wi-Fi technologies continues to grow, coordination beyond LBT will become necessary to maintain fairness and provide high quality of experience. A local inter-technology channel will play a strong role in the effectiveness of MAC protocols and coordination.

| uLTE Parameter | Wi-Fi Beacon Field | Purpose |
| --- | --- | --- |
| Operator / Cell ID / PLMN | SSID | Identification |
| Channel number | DS Parameter Set | Identification+ Channel Selection |
| Channel number | HT operation | Identification |
| Loading: Station count, Channel Utilization, Available Admission Capability | BSS Load | Load Balancing / Admission Control |
| Node Type: { Rel-X LAA / MulteFire / LTE-U } | Vendor Specific Field | Identification+ Channel Selection |
| MAC Spec: {LBT Cat-X, other } | Vendor Specific Field | Identification |
| TX power offset (relative to AP beacon) in dBm | Vendor Specific Field | Channel Selection |

Table 2. Example of 802.11 beacon fields populated with *u*LTE data.